# From Digital Accountability to Accountable Digitality Through Needs-Aware Information Systems: The Case of Auditable Child-Welfare Judgments

*Completed Research Paper*

**Soheil Human**
Vienna University of Economics and Business, Vienna, Austria
IT:U Interdisciplinary Transformation University Austria, Linz, Austria
Delft University of Technology, Delft, The Netherlands
soheil.human@wu.ac.at

## Abstract

*Digital accountability research asks how digital systems can, among other aims, be made transparent, explainable, auditable, contestable, and supportive of ongoing learning and improvement. This paper reverses the question: how can digital transformation make established human institutions more accountable? It theorizes this reversal as accountable digitality and specifies needs-aware information systems as the mediating mechanism. The hard and paradigmatic case is child-welfare judgment, where best-interest procedures must protect children, preserve confidentiality, and respect judicial independence while enabling aggregate learning about needs, reasons, exceptions, and disparities. The case is used diagnostically and illustratively to derive and examine the design logic, not as empirical evidence or validation. Conceptual design-oriented analysis decomposes and recombines digital and legal accountability under child-rights constraints, deriving a canonical theory-to-design chain, contingent mechanisms, implications, and safeguards. It advances IS responsibility and ethics research by showing how privacy-preserving, co-created, needs-aware information systems can support institutional self-knowledge and auditable justice.*

## Introduction

Digital accountability has become a central problem for information systems (IS) research because digital systems increasingly participate in consequential institutional decisions. Early public computerization already raised concerns that databases and administrative information systems could turn social categories into institutional facts while hiding how classifications were produced and how they might be contested (Friedman & Nissenbaum, 1996; Kling, 1991). Machine learning intensified this problem because predictive systems can be opaque, reproduce historical bias, and distribute responsibility across data producers, model developers, organizational deployers, users, and regulators (Barocas & Selbst, 2016; Burrell, 2016; Kroll et al., 2017). Foundation models and generative artificial intelligence (GAI) have further expanded the accountability problem into knowledge work and professional advice, where outputs may be persuasive, difficult to verify, and consequential even when formally advisory (Bender et al., 2021; Berente et al., 2021).

The dominant direction of this debate asks how digital systems can be made accountable. This paper develops the reverse and complementary question: how can concepts developed for *digital accountability* make established human institutions more accountable once they become mediated by information systems? The question matters because many high-impact decisions remain formally human and legal while exhibiting accountability problems similar to those that motivate algorithmic accountability research:

opacity, uneven documentation, limited contestability, difficulty detecting bias, and a weak connection between individual decisions and aggregate institutional patterns (Bovens, 2007; Koppell, 2005; Mashaw, 2006; Power, 1997). The asymmetry is striking. If an AI system supported a judicial decision, it would normally be expected to have documentation, testing, monitoring, and explanation. Yet the aggregate behavior of human legal decision-making systems can remain difficult to observe because cases are dispersed, confidential, local, and highly contextual.

This reversal is theorized as a movement from *digital accountability* to *accountable digitality*. Zuboff's (1988) distinction between automation and "informating"—the capacity of digital technologies to generate information about activities they mediate—provides historical context. Accountable digitality develops this visibility-producing potential toward institutional answerability, but the concepts are not equivalent: visibility becomes accountability only when linked to reasons, safeguards, contestability, authorized forums, and learning. Digital accountability concerns the answerability of digital systems and their organizational operators. Accountable digitality concerns the use of *digital transformation* itself to increase the answerability of established institutions without collapsing their normative commitments into computational control. Information systems can thus become institutional observatories that document reasons, preserve privacy, surface aggregate patterns, and furnish authorized forums with evidence for questioning and evaluating institutional conduct (Bovens, 2007; Koppell, 2005; Mashaw, 2006). The proposed construct is not an argument for automated justice. It is an argument for auditable justice.

The theoretical case is child-welfare judgment, especially best-interest assessment in post-separation contexts. This hard and paradigmatic case combines rights, science, confidentiality, judicial independence, discretion, and profound consequences while exposing tensions shared by other domains of dispersed institutional decision-making. Here, the focal legal information system is the socio-technical arrangement of people, rules, records, routines, digital infrastructure, and accountability forums through which cases are represented, reasoned, decided, reviewed, and learned from—not merely a software application or court case-management system. A *needs-aware information system* represents human needs as contestable normative-empirical constructs. At the case level, its unit of analysis is the *Need–Satisfier Linkage* and *Reasoning Record* (shortened hereafter to *needs-aware reasoning record*): a traceable and contestable record linking identified or hypothesized needs to candidate satisfiers and alternatives, evidence, reasons, stakeholder perspectives, uncertainty, trade-offs, and the decision, while keeping any later evidence of *need satisfaction* separate. Protected aggregation makes patterns across such records reviewable. Its core mechanism is governed, reasoned linkage and review rather than reliance on need scoring or automated outcome determination. Child welfare is used as a demanding running case because the gap between individual judgment and aggregate accountability is especially visible when decisions shape children's developmental contexts, not because accountable digitality is confined to this domain.

The research is guided by two questions: *How can concepts and methods from digital accountability be translated into needs-aware information systems that strengthen the accountability of established human institutions while preserving context, rights, and institutional responsibility? How can this general design logic be instantiated in child-welfare judgment while protecting children's rights, privacy, confidentiality, and judicial independence?*

The paper makes three contributions to IS research. First, it conceptualizes *accountable digitality* as a theoretical move: *digital transformation* can be designed to account for institutions, not only to be accounted for by them. Second, it develops *needs-aware information systems* as the mechanism through which accountable digitality becomes substantive rather than procedural. Third, it specifies a privacy-preserving and independence-preserving audit model for child-welfare judgment, including general meta-requirements, child-welfare instantiations, candidate architecture realizations, contingent mechanisms, risks, and safeguards. The contribution is *conceptual* and *design-oriented*: it explains what must be made visible, why visibility must be limited, and how IS can support institutional accountability without substituting computational metrics for legal judgment (Gregor, 2006; Hevner et al., 2004; Rowe, 2012).

## Research Approach and Contribution Type

This paper is a *conceptual theory-building* article with a *design-oriented output*. It develops the construct of accountable digitality and nascent design knowledge in the form of a problem class, general meta-requirements, child-welfare instantiations, candidate architecture realizations, contingent mechanisms, safeguards, and boundary conditions. Design-science concepts are used to articulate these outputs and guide future design and evaluation; no implemented artifact or completed design-science project is claimed. The child-welfare material is used diagnostically and illustratively to develop and examine the design logic, not as an empirical case study or validation. In IS terms, a *theory for analysis and explanation* with implications for socio-technical accountability infrastructures is contributed (Gregor, 2006; Hevner et al., 2004; Rowe, 2012). Rigor is assessed through construct clarity, transparent derivation, theoretical usefulness, mechanism specification, and explicit boundary conditions (Weber, 2012; Whetten, 1989).

### *Stepwise Conceptual Procedure*

The conceptual procedure follows five analytical moves. First, it problematizes the usual direction of accountability by asking whether digital accountability concepts can make established human institutions more accountable once they are mediated by information systems (Alvesson & Sandberg, 2011; Gupta, 2018; Rai, 2017a, 2017b). Second, it decomposes digital accountability into documentation, transparency, explanation, audit, contestability, monitoring, and governance, and decomposes child-welfare judgment into independence, confidentiality, appeal, reason-giving, child participation, best-interest procedure, and institutional learning. Third, it recombines these elements with child-rights, human-needs, privacy, and judicial-independence constraints to define accountable digitality and needs-aware information systems. Fourth, it specifies child-welfare judgment as a paradigmatic case because it exposes the tension between individualized confidential judgment and aggregate accountability (Flyvbjerg, 2006). Fifth, it derives general meta-requirements, child-welfare instantiations, candidate architecture realizations, contingent mechanisms, implications, risks, and safeguards. Together, these moves form a bounded theory-to-design chain across general and child-welfare levels (Jaakkola, 2020; Whetten, 1989).

The design-problem class comprises repeated and distributed institutional decisions, practices, or service outputs whose individual instances may appear defensible while aggregate patterns remain difficult to observe, explain, contest, or improve under confidentiality, discretion, unequal power, or resource constraints. The bounded design knowledge offered is a general accountability logic that connects theoretical warrants and general meta-requirements to needs-aware records, protected aggregate observability, authorized forums, safeguards, and institutional learning. The child-welfare architecture is the hard, paradigmatic instantiation through which that logic is developed; it neither confines the construct to child welfare nor establishes domain-free validity. Transfer requires each domain's purposes, needs, rights, risks, data, and governance to be specified anew.

## Digital Accountability as a Source Domain

### *A Brief Historical Trajectory*

Concerns about accountability in digital decision environments did not begin with contemporary AI. Public computerization and administrative databases already changed how organizations classified people, allocated services, and treated records as institutional facts. Early IS and science-and-technology scholarship therefore warned that computerization could redistribute power, make categories durable, and hide value choices inside apparently technical infrastructures (Friedman & Nissenbaum, 1996; Kling, 1991). In this first era, accountability questions centered on access, data quality, classification, organizational control, and the possibility that administrative information systems could make persons legible to institutions while making institutions less legible to persons.

The machine-learning era intensified these concerns because decision rules could be inferred from data rather than written as explicit organizational procedures. Bias could enter through historical data, target definitions, sampling, labels, feature construction, and deployment contexts; opacity could arise from model complexity, trade secrecy, organizational fragmentation, or the gap between statistical explanation and human justification (Barocas & Selbst, 2016; Burrell, 2016; Kroll et al., 2017). The accountability object therefore expanded from software correctness to sociotechnical decision systems: data, models, institutional purposes, implementation settings, and recourse mechanisms had to be considered together.

Foundation models, generative AI, and more agentic forms of AI introduce a further shift, but the shift should not be overstated for this paper. These systems can produce text, recommendations, summaries, or action proposals that appear context-sensitive and authoritative; they can be embedded in workflows as assistants, monitors, or semi-autonomous components; and they can make responsibility harder to allocate when outputs emerge from distributed development, fine-tuning, prompting, and organizational use (Bender et al., 2021; Berente et al., 2021). Their relevance here is instrumental. They are possible components of needs-aware information systems, not the central theoretical object.

Across these eras, the accountability repertoire has become more concrete. Documentation instruments such as datasheets and model cards, internal audit frameworks, interpretability methods, fairness analysis, and post-deployment monitoring have created a vocabulary for asking what is known, what is documented, who can challenge outputs, and how harms can be detected (Doshi-Velez & Kim, 2017; Gebru et al., 2021; Mitchell et al., 2019; Raji et al., 2020). *Accountable digitality* asks whether this repertoire can be translated to institutions that are not themselves algorithms but that increasingly depend on digital records, digital workflows, and digital forms of institutional memory.

### *Digital Accountability Mechanisms and Institutional Learning*

Accountability is a relational concept. An actor is accountable when the actor must provide an account of conduct to a forum that can question, evaluate, and impose consequences (Bovens, 2007). Public administration and legal scholarship distinguish hierarchical, legal, professional, political, and social accountability, each with different forums and standards of evaluation (Koppell, 2005; Mashaw, 2006). IS research extends this relational view because digital infrastructures distribute action across systems, organizations, routines, and data practices (Kling, 1991; Orlikowski, 1992; Roberts, 1991).

Algorithmic and AI accountability research offers transparency, explainability, documentation, audits, model cards, datasheets, fairness metrics, human oversight, recourse, and post-deployment monitoring (Diakopoulos, 2015; Doshi-Velez & Kim, 2017; Gebru et al., 2021; Mitchell et al., 2019; Raji et al., 2020). Their value, however, is bounded: transparency can yield visibility without understanding; explanation can rationalize outputs without locating institutional responsibility; fairness metrics can conflict; and audits can become symbolic when forums lack authority, expertise, or access (Ananny & Crawford, 2018; Power, 1997; Selbst et al., 2019). Thus, accountability requires institutional design rather than reliance on a transparent algorithm for every consequential decision: appropriate forums can examine reasons, roles, data, and consequences, and findings can prompt contestation, reasoned response, and learning rather than disclosure alone. This extends beyond algorithmic systems. Here, legal institutions are socio-technical information systems that classify cases, record claims, transform evidence into formal categories, route files among actors, and produce authoritative outputs (Kling, 1991; Orlikowski, 1992). Court files are not mere archives but structured representations of children, parents, harms, needs, facts, procedural histories, and legal options. Digital transformation changes what can be observed across them, by whom, and why; it can reveal patterns obscured by paper-based or siloed systems but also subject institutional judgment to targets, dashboards, or decontextualized comparisons. Accountable digitality therefore draws on digital accountability as a *source domain* with safeguards against *technological solutionism*.

### *From Accountable Digital Systems to Accountable Institutions*

The move from accountable digital systems to accountable institutions reframes the question of bias. In algorithmic settings, bias is often investigated through training data, model behavior, output distributions, and organizational deployment (Barocas & Selbst, 2016; Friedman & Nissenbaum, 1996). In legal settings, bias may arise from human cognition, professional routines, evidentiary asymmetries, cultural assumptions, regional norms, workload, and institutional histories. Research shows that judges, like other humans, can be affected by heuristics, anchoring, implicit associations, and contextual information (Englich et al., 2006; Guthrie et al., 2001; Rachlinski et al., 2009). Judicial independence is a crucial safeguard against external pressure, not a guarantee against internal or systemic bias. This distinction matters for audit design. Publicly ranking individual judges on raw outcomes may create metric pressure, invite strategic behavior, and obscure case mix (Council of Europe, Committee of Ministers, 2010; Power, 1997; Selbst et al., 2019). By contrast, an arrangement organized around *needs-aware reasoning records* and protected aggregation is intended to support *institutional learning* without individual scorecards. The accountable forum may be an internal judicial training body, an appellate court, a child-rights oversight institution, a data-protection authority, or a carefully governed public reporting mechanism. Accountable digitality therefore requires multiple accountability forums and visibility levels, rather than a single transparency ideal.

## Needs-Aware Information Systems

A *needs-aware information system* is a socio-technical arrangement that represents [human] *needs* as contestable normative-empirical constructs, links decisions to those needs, records uncertainty and exceptions, enables privacy-preserving audit, and supports forums of contestation and learning. The concept builds on needs-aware AI research, which argues that AI and digital systems should not optimize only for preferences, engagement, utility, or satisfaction because these proxies can diverge from human flourishing (Human & Watkins, 2023; Watkins & Human, 2023). It also builds on knowledge representation work that treats [human] *needs* and *satisfiers* as concepts that can be modeled, debated, and refined (Human et al., 2017).

Needs-awareness is not paternalism when properly designed. A *need* may be implicit, hypothesized, contextual, or contested; a *satisfier* is a context-specific object, state, action, activity, service, environment, or arrangement that may contribute to meeting it. The relationship is many-to-many. At decision time, *satisfiers* are candidates and their proposed links to *needs* require evidence and reasons. *Need satisfaction* refers only to the later process or evidence that a *need* was actually met and cannot be inferred from selecting a candidate *satisfier* (Doyal & Gough, 1991; Human et al., 2017; Human & Watkins, 2023; Watkins & Human, 2023). *Needs* are therefore structured questions, not rigid commands that determine outcomes. For children, they must be interpreted in relation to development, vulnerability, dependency, participation, and ecological, cultural, social, and identity contexts, including experiences of marginalization (Bronfenbrenner, 1979; Lundy, 2007). A needs-aware system should record general baselines and case-specific reasons for departure.

Needs-awareness makes accountable digitality substantive. A transparent decision that merely states an outcome is not enough. An explainable AI component that summarizes documents is not enough. An audit dashboard that counts cases is not enough. Accountability becomes meaningful when the institution can show how its decisions relate to the child's safety, emotional security, attachments, identity, language, culture, education, health, participation, and future development. This is why needs-aware IS are especially relevant for child-welfare judgment: they connect rights, empirical knowledge, and institutional data without treating any one of them as sufficient by itself.

The International Organization for Standardization (ISO) established Technical Committee 356 on children's rights management in 2026, signaling an emerging effort to support the implementation of children's protection, provision, and participation rights under the Convention on the Rights of the Child (International Organization for Standardization, n.d.). Such standardization can support more consistent and rights-sensitive best-interest procedures and should also address how information systems can document and audit confidential decisions without exposing children or improperly constraining lawful discretion. This emerging agenda underscores the timeliness and practical relevance of accountable digitality and needs-aware information systems.

Needs-awareness is especially important because rights can be too abstract for audit and data can be too thin for justice. A right to family life, participation, education, cultural identity, or protection from harm must be translated into questions that a decision process can answer without reducing the child to a checklist. Needs categories provide this middle layer. They identify what the system must attend to—safety, attachments, stability, participation, education, cultural continuity, health, and holistic development over time—while leaving room for legal judgment about how these needs interact in a particular case.

A needs-aware information system should therefore be designed around defeasible attention rather than deterministic scoring. The system should ask whether a need was considered, what evidence was used, how the child's views were heard, why a particular arrangement was selected, and why alternative arrangements were rejected. This is different from assigning a numerical best-interest score. It is also different from merely storing documents. The design goal is to make the reasoning chain inspectable enough for institutional learning while preserving the legal and ethical fact that children are rights-holders whose lives cannot be optimized as if they were production functions.

This distinction also clarifies the role of child participation. Needs-awareness is not a paternalistic substitute for voice. Article 12 of the Convention on the Rights of the Child requires that children be heard in matters affecting them, and child-rights scholarship shows that voice must be supported by space, audience, influence, and information rather than treated as a symbolic procedural step (Lundy, 2007; United Nations, 1989). A needs-aware system should therefore document not only adult assessments of needs but also how the child's perspective was elicited, understood, weighed, and protected from manipulation or burden.

## Child-Welfare Judgments as a Hard and Paradigmatic Case

### *Child Rights Assessment, Best-Interest Assessment, and Needs*

The best interests of the child are a central principle of the United Nations Convention on the Rights of the Child (UNCRC). Article 3 requires that the best interests of the child be a primary consideration in all actions concerning children, and the Committee on the Rights of the Child interprets this principle as a substantive right, an interpretive principle, and a rule of procedure (United Nations, 1989; United Nations Committee on the Rights of the Child, 2013). The procedural dimension is crucial for IS: a best-interest justification must identify the child's relevant interests, criteria, factual circumstances, how those interests were weighed, and how the child's *needs* were taken into account (United Nations Committee on the Rights of the Child, 2013). This procedural adequacy is a condition for normative assessment; it is not proof that an outcome is correct or that an institution is sociologically legitimate.

*Child-rights impact assessment* examines effects on children collectively at the policy or system level (Payne, 2019), whereas *best-interest assessment* concerns the circumstances of a particular child (United Nations Committee on the Rights of the Child, 2013). Accountable digitality connects these analytically distinct levels. Individual decisions generate protected institutional knowledge when their reasons are documented in comparable ways; aggregate child-rights assessment becomes more precise when it can learn from patterns across individual decisions. The theoretical importance of the child-welfare case lies in this two-way movement between the individual and the institutional.

*Needs* provide the bridge between these levels. The child's best interests cannot be reduced to parental claims, legal categories, or administrative convenience. They require attention to safety, health, stable care, attachments, participation, identity, education, cultures, languages, extended family, and holistic development over time. Some of these needs are universal at a high level; their satisfiers may vary by child, family, community, and circumstance. A needs-aware information system should therefore not impose a single outcome. It should require that relevant needs be considered, that departures from general baselines be reasoned, and that institutional patterns be learnable.

### *Why the Case Is Hard and Paradigmatic*

The case is hard because the institution must reason across several forms of knowledge that do not collapse into one another. Law supplies rights, procedural duties, presumptions, and standards of proof. Cognitive Science, psychology, neuroscience, linguistics, sociology, education, and family studies supply empirical knowledge about development, attachment, conflict, resilience, and harm. Children and families supply situated knowledge about relationships, fear, care, languages, cultures, and practical feasibility. Judges must integrate these forms of knowledge without turning any single one into a mechanical rule. The difficulty is therefore not a reason to avoid audit. It is the reason audit must focus on reasons, distributions, and safeguards rather than on outcomes alone.

The case is paradigmatic because confidentiality and discretion are often treated as if they exhaust the accountability problem. They do not. Confidentiality explains why child-welfare data cannot be openly published. Discretion explains why each case must be reasoned individually. Neither explains why an institution should be unable to know whether its decisions systematically neglect some needs, whether children in comparable situations receive radically different developmental contexts, or whether particular reasons are invoked unevenly across gender, class, culture, language, disability, or region. The paradigmatic lesson is that *aggregate accountability* is not the enemy of individualized justice; it is one of the conditions under which individualized justice can be evaluated.

### *Case-by-Case Reasoning Is Not Equal-Validity Epistemic Relativism*

Best-interest assessment is often described as case by case, and rightly so. A child's safety, voice, attachments, health, schooling, cultural identities, and family relationships cannot be determined by a universal formula. Yet case-by-case reasoning does not imply that form of epistemic relativism which treats all claims as *equally valid* ("everything goes"). No one would infer from individualization that one child does not need water, sleep, safety, or emotional security. The more difficult question is not whether children have needs but how decision-making systems reason about needs when evidence is complex, contested, and institutionally dispersed.

This point is particularly important in post-separation decisions about where children live and how their relationships with each parent are supported. Empirical research suggests that *shared physical custody* and substantial involvement of both parents can be associated with favorable child outcomes in many circumstances, while also emphasizing selection effects, conflict, safety, and contextual conditions (Bauserman, 2002; Bergström et al., 2015; Steinbach, 2019). The evidence does not create a universal rule that equal time is always best. It does create a baseline question: when a child's developmental context is narrowed to one parent as primary home and the other as visitor, what *needs-based reasons* justify that arrangement, and how are such reasons distributed across cases? A *needs-aware audit* asks this question without replacing judicial judgment.

Comparative data make this institutional pattern visible. Using weighted 2021 data from 17 European countries, Hakovirta et al. (2023) report that 42.5% of children in separated families in Sweden lived in *equal joint physical custody*, whereas Austria was among nine countries in which the corresponding share was 5% or less. This contrast does not, by itself, establish which approach better serves children; differences in law, culture, housing, parental work patterns, and safety require careful interpretation. It nevertheless reveals a consequential institutional reality: children in broadly comparable European welfare states are being placed in markedly different developmental contexts. Accountable institutions should therefore be able to examine and explain whether such patterned differences reflect children's needs and circumstances, legal standards, case mix, social constraints, or institutional bias.

### *The Bicultural Child as a Diagnostic Example*

The bicultural child illustrates why aggregate auditability matters. When parents come from different linguistic, cultural, religious, or national communities, post-separation arrangements can shape identity formation, language acquisition, relationships with extended family, and access to cultural practices (Bronfenbrenner, 1979; United Nations, 1989; United Nations Committee on the Rights of the Child, 2013). A single outcome may be justified by safety, distance, the child's views, or practical constraints. Across many cases, however, repeated patterns may reveal that one side of a child's heritage is systematically marginalized. Without aggregate data, a child-rights forum cannot ask whether intercultural children receive arrangements that preserve language and identity needs, whether fathers or mothers are routinely positioned as visitors, or whether children's expressed views are documented consistently.

Read analytically, not as evidence, the child-welfare case generates the requirements. *Contestable developmental needs* require substantive representation; *confidentiality and judicial independence* require calibrated visibility; *dispersed judgments* require protected cross-case observability; *detected patterns* require authorized forums and response; *digital mediation* requires validation, monitoring, and contestation; and *children and family members* require safeguarded participation. These concerns inform six general meta-requirements (MR1–6); the case motivates their derivation without empirical validation.

## Accountability Gaps in the Current Legal Information System

### *Independence, Layering, and Confidentiality*

Legal systems contain important accountability safeguards, although their form varies by jurisdiction. Judicial independence protects adjudication from improper external and internal influence; reasoned judgments and appeal create forums for reviewing reasons, evidence, and procedure; and confidentiality protects children and families from exposure (Council of Europe, Committee of Ministers, 2010; Mashaw, 2006; United Nations, 1989). These safeguards matter because independence and confidentiality protect rights that an accountability infrastructure must not displace.

These safeguards do not eliminate cross-case blind spots. Judicial independence is not a guarantee against cognitive bias, local norms, workload, stereotypes, or incomplete knowledge (Englich et al., 2006; Guthrie et al., 2001; Rachlinski et al., 2009). Because appeal reviews only cases that enter the appellate process, it cannot by itself reveal routine cross-case patterns. Confidentiality protects privacy but can restrict institutional visibility. The result may be a system that is accountable in formal legal terms but weakly auditable as a distributed information system.

This is not a failure of individual judges alone. It is an architectural problem. If case files do not use needs-aware reasoning records, if local courts cannot compare patterns, if appellate courts see only selected cases, and if policymakers lack privacy-preserving aggregate indicators, then the system cannot easily learn whether it is producing child-centered decisions at scale. *Accountable digitality* therefore treats the legal system as a *distributed socio-technical system* whose accountability depends on documentation, data governance, and learning loops, not only on individual integrity.

### *The Digital Paradox of Human Judgment*

A paradox follows. Contemporary society increasingly expects transparency, explainability, fairness analysis, and auditability from AI systems, particularly in high-stakes settings. Regulatory and scholarly debates around high-risk AI systems emphasize documentation, risk management, oversight, and post-deployment monitoring (European Parliament & Council of the European Union, 2024; Raji et al., 2020). Yet the human legal systems that AI might support are often much less auditable in aggregate. A human judge may write reasons in a single case, but the institution may not know how comparable cases are reasoned across courts, regions, cultures, parental categories, or child needs.

This paradox does not imply that AI is more legitimate than judicial judgment. It implies that *digital accountability* has raised the *accountability expectations* for any consequential decision system. If an AI component would be considered unacceptable without documentation, audit, and contestability, then a human institution should not be insulated from all *aggregate accountability* merely because decisions are human. The appropriate response is not automated judging; it is *accountable digitality*: information systems that help institutions become more reflective, more evidence-sensitive, and more answerable while preserving the normative role of human legal judgment.

### *What the Legal Information System Fails to Know*

A legal information system can answer case-management questions yet remain unable to support accountability questions. Recording filing dates, hearings, parties, procedural steps, and closure is insufficient if the system cannot represent *needs*, candidate *satisfiers*, reasons, rejected alternatives, child participation, protected cross-case patterns, or later evidence of *needs satisfaction*. This is the institutional knowledge gap addressed here: a system may process cases without being able to examine whether its procedures are becoming more child-centered over time.

Consider post-separation arrangements. A court may decide how much time a child spends in each developmental context, but an oversight body may not be able to ask, in a privacy-preserving way, how often children are placed in an *equal two-parent developmental context*, how often one parent is reduced to visitor-like *contact*, how these patterns differ by region, age, culture, disability, conflict allegations, or language, and what reasons are given for deviations. Comparative and family research shows that *shared physical custody* is uneven across Europe and can be associated with child wellbeing under appropriate conditions, while also requiring attention to conflict, resources, and safety (Bauserman, 2002; Bergström et al., 2015; Hakovirta et al., 2023; Steinbach, 2019).

The accountability point does not depend on a universal custody preference. It depends on the institution's ability to distinguish justified exceptions from unexplained patterns. If children with intercultural parents routinely lose contact with one language community, if fathers or mothers are systematically constructed as visitors without case-specific safety reasons, if children's stated views disappear from reasoning, or if comparable courts produce very different arrangements without explanation, the institution should be able to see these patterns. Without *needs-aware data structures*, the system lacks the informational conditions for asking such questions responsibly.

Publication of appellate judgments cannot by itself reveal routine institutional practice because it captures only cases that reach and are selected for appellate review. Such judgments are important for doctrine but incomplete as institutional sensors. *Accountable digitality* supplements doctrine with protected aggregate observability. It asks not only what the law says in leading cases, but how the legal information system behaves across the everyday cases in which children's developmental contexts are configured.

## A Needs-Aware Architecture for Auditable Child-Welfare Judgment

Building on the earlier analysis of *accountable digitality*, *needs-aware information systems*, and the child-welfare accountability gap, this section translates the theoretical argument into a conceptual sociotechnical architecture. It specifies a governed pathway from needs-aware reasoning records and purpose-limited visibility through protected cross-case inquiry to authorized response and learning, with child rights, privacy, participation, and judicial independence governing each step. This design logic is not a software blueprint, scoring model, or automated decision system. Six general *meta-requirements* address how distributed institutional conduct can become observable and answerable without displacing contextual judgment or responsibility. Child-welfare instantiations, staged candidate capabilities, safeguards, responsible actors, and evaluative questions show how this logic could be realized and examined in the hard and paradigmatic case.

### General Meta-Requirements and Child-Welfare Instantiations

Five warrants jointly inform the meta-requirements; none suffices alone. *Relational accountability* requires records, reasons, authorized forums, questioning, evaluation, and response (Bovens, 2007; Mashaw, 2006). *Child rights* require best-interest procedure, participation, identity, protection, and privacy (United Nations, 1989; United Nations Committee on the Rights of the Child, 2013). *Needs theory* requires contestable links among needs, candidate satisfiers, evidence, and reasons without treating selection as need satisfaction (Doyal & Gough, 1991; Human et al., 2017; Human & Watkins, 2023); the capability approach similarly distinguishes available means from achieved well-being (Sen, 1999). *Contextual privacy* requires purpose-limited, role-sensitive visibility (Nissenbaum, 2004). *Judicial independence* protects adjudication from improper influence while permitting reason-giving, lawful review, and institutional learning (Council of Europe, Committee of Ministers, 2010). Relational accountability, needs theory, and contextual privacy are transferable across settings; child rights and judicial independence add constraints specific to the focal adjudicative setting. Other domains require renewed normative and institutional justification. Needs-aware categories supplement rather than displace statutory criteria, evidentiary requirements, or rights. The architecture therefore pursues accountable representation and reasoning. In protected cross-case inquiry, sensitive attributes may be analyzed only when lawful, necessary, purpose-limited, and methodologically justified; they must not determine outcomes or rankings, and associations alone establish neither bias nor causation.

### Staged Capabilities and Theory-to-Design Mapping

*Stages* mark proposed capability dependencies, not validated or mandatory implementation phases: Stage 1 establishes governed case records; Stage 2 builds protected cross-case inquiry and response on them; optional Stage 3 adds bounded AI assistance. Stage 1 combines a contestable vocabulary of needs and candidate satisfiers with structured reasoning records, minimization, role-based access, identifier separation, provenance and access logs, correction, and case-level challenge. Correcting information does not alter a judgment; applicable review or appeal procedures govern any such change. Stage 2 adds, where appropriate, pseudonymized authorized review, protected aggregates, denominators, missingness and contextual variables, sampling, methodological and access audits, monitoring, and documented response. Federated methods may reduce raw-data movement but neither guarantee privacy nor constitute an architectural requirement (Kairouz et al., 2021). Stage 3 permits only bounded field-presence checks, source-linked summaries, or review prioritization after task-specific validation, documented limitations, human verification, challenge and correction, error and drift monitoring, and provision for suspension or disablement; it neither evaluates normative reason quality nor recommends or determines outcomes (Bender et al., 2021; Raji et al., 2020). Governing parameters include vocabulary versions, required and optional fields, access roles, retention and correction rules, aggregation and small-count thresholds, contextualization rules, reporting cadence, response responsibilities, and, for AI, validation criteria and incident procedures.

Table 1 consolidates the many-to-many mapping from five warrants through six meta-requirements to child-welfare realizations, stages, risks, safeguards, governance responsibilities, and evaluative questions. Warrants are principal rather than exclusive, and responsibilities remain jurisdiction-dependent. MR1 concerns represented content; MR2, access; MR3, cross-case inquiry; MR4, response and learning; MR5, sociotechnical mediation; and MR6, participation and challenge. The table synthesizes nascent design knowledge, not a validated architecture or deterministic derivation.

### From Needs-Aware Reasoning Records to Governed Audit and Learning

Together, the meta-requirements suggest four contingent mechanisms. Needs-aware reasoning records may improve answerability by making need–candidate-satisfier linkages, evidence, alternatives, uncertainty, and exceptions inspectable. Protected aggregation may surface contextual signals of omissions or disparities. Authorized forums may translate signals into questioning, contestation, response, redress, or learning. Finally, inspectable and contestable sociotechnical mediation—including optional AI—may support record completeness and authorized review without displacing human answerability, provided it remains purpose-limited, validated, monitored, correctable, and subordinate to responsible judgment.

Applied analytically, the architecture traces contestable needs and candidate satisfiers from case records under restricted access through protected cross-case inquiry to authorized contextual review, response, and learning; it neither simulates nor prescribes a best-interest outcome, and judgment remains with responsible institutional actors. It supports questions such as: Which needs, candidate satisfiers, and feasible alternatives were considered, and what evidence and uncertainty supported each linkage? Was the child enabled to express views accessibly, age-appropriately, safely, and without coercion, with those views documented and given due weight according to age and maturity? Were relationships with each parent and relevant caregiver, caregiving histories, developmental significance, safety, and feasibility assessed individually without gendered presumptions alongside sibling and extended-family relationships, identity, cultural and linguistic continuity, school stability, and other developmental conditions? Were departures and exceptions explained? Across comparable cases, were materially similar considerations addressed consistently while preserving legal and contextual differences and reasons for different treatment? These questions examine reasoning; they neither establish *need satisfaction* nor operate as decision rules.

**Table 1. Theory-to-Design Mapping for Accountable Digitality in Child Welfare**

| **Meta-requirements and warrants** | **Child-welfare instantiation → staged realization** | **Principal risk → safeguard → responsible actor/locus** | **Bounded question for later evaluation** |
|---|---|---|---|
| **MR1 Substantive and contestable representation** — accountability; child rights; needs | Needs-aware reasoning record → Stage 1 contestable vocabulary, structured capture, provenance, and correction/contestation; no outcome or need-satisfaction inference. | Frozen categories or ritualization → periodic vocabulary and sampled-reason review → child-rights/professional forum; vocabulary and record loci. | Does record presence and quality improve without narrowing judgment? |
| **MR2 Calibrated and purpose-limited visibility** — accountability; privacy; child rights; independence | Identifiable case material, separated or pseudonymized local review, and protected aggregates → Stages 1–2 minimization, role-based access, access logs, suppression, and thresholds. | Reidentification or surveillance → least privilege and purpose limits → data steward/privacy authority; access and aggregation loci. | Can lawful scrutiny occur without unnecessary exposure or judge ranking? |
| **MR3 Contextual distributional observability** — accountability; needs; child rights; privacy | Needs, reasons, omissions, and decisions contextualized by safety, age, care history, feasibility, child views, and law → Stage 2 reporting with denominators, missingness, case mix, uncertainty, and authorized sampling. | Metric fixation or false inference → inquiry prompts, qualitative follow-up, and no performance scores → authorized audit forum; reporting locus. | Do protected signals identify patterns warranting contextual review without treating observed disparities alone as conclusive evidence of bias or causation? |
| **MR4 Authorized forum, response, and learning** — accountability; child rights; independence | Appellate, judicial-learning, child-rights, privacy, or other lawful forums → Stage 2 issue routing, sampled-case review, response log, and remediation/learning monitoring. | Symbolic audit, overreach, or unequal capacity → lawful mandate, proportional inquiry, staged minimum, and resourcing review → institutional leadership/oversight; response locus. | Do signals produce reasoned action, learning, or redress without controlling judgments? |
| **MR5 Accountable sociotechnical mediation** — accountability; privacy; independence | Inspectable templates, analytics, dashboards, and optional AI → Stages 1–3 documentation, validation, human review, monitoring, correction, and disablement; no outcome determination or ranking. | Error, automation bias, or opacity → bounded tasks, validation, incident handling, and override → system owner/assurance body; component lifecycle. | Does assistance improve completeness or audit selection without degrading rights or responsibility? |
| **MR6 Affected-person participation and rights-sensitive representation** — child rights; accountability; needs; privacy | Child voice and party challenges in records and governance → Stage 1 accessible explanation, participation metadata, and correction routes; Stage 2 input to indicator review. | Tokenism, coercion, or unsafe disclosure → age-appropriate representation, safety exceptions, and recorded response → child representative/case forum; intake, record, and governance loci. | Are affected perspectives heard and acted upon without unsafe exposure? |

To connect case-level reasons to institutional audit, this paper develops three nonscalar analytical dimensions. *Reason presence* asks whether the record contains, where applicable, an identified *need*, candidate *satisfiers*, evidence, alternatives, uncertainty, and any departure with its stated justification; it concerns completeness. *Reason quality* asks whether a need–candidate-satisfier linkage is specific, evidence-linked, uncertainty-aware, responsive to plausible alternatives, and contestable; it concerns answerability (Bovens, 2007; Mashaw, 2006; United Nations Committee on the Rights of the Child, 2013). *Reason pattern* asks how *needs*, candidate *satisfiers*, reasons, omissions, exceptions, and decisions are distributed across comparable cases; it is a system-level diagnostic and does not itself establish bias, discrimination, or causation (Barocas & Selbst, 2016; Selbst et al., 2019). Structured records enable presence review, authorized human forums assess quality, and protected aggregation surfaces patterns for inquiry. AI may assist with bounded presence checks, source-linked summaries, or flags, but it is not treated as establishing legal adequacy, factual correctness, or substantive justification.

Governance assigns purposes, authority, access, methods, response, and revision; privacy constrains information flows; transparency makes relevant records, methods, and limits visible; explanation supplies role-relevant reasons, provenance, uncertainty, and limits; audit examines records and patterns; contestability enables correction and challenge; and monitoring tests data, components, and effects over time. Visibility and explanation are role-specific, but roles alone do not justify access: each flow requires a lawful purpose, necessity and proportionality, and the least identifying form sufficient for its function. Affected persons and representatives receive case-relevant reasons, provenance, uncertainty, and separate routes to correct records or challenge judgments. Correction does not change a judgment; applicable review or appeal governs any such change. Professionals receive provenance, definitions, method limits, and the basis of flags. Authorized oversight receives disclosure-controlled aggregates and, where mandate and necessity require, sampled records under enhanced controls. Public audiences may receive lawful and safe disclosure-controlled trends, methods, caveats, and responses, never identifiable child or family information. No judge-level or other individual ranking is authorized. Effects on professional acceptance, affected-person trust, or engagement remain empirical.

Implementation requires jurisdiction-appropriate, safeguarded co-design. Judges, child representatives, psychologists, data stewards, and oversight bodies would help define relevant *needs*, sufficient reasons, and justified exceptions; children and families would participate through safeguarded channels. Yet participation does not neutralize power: records could support inquiry or be repurposed by court leadership, ministries, vendors, or professional bodies for surveillance, discipline, workload control, or public comparison (Kellogg et al., 2020; Zuboff, 1988). Uses should therefore be purpose-limited, access and secondary use logged, and the learning layer separated from individual ranking; affected persons and professionals should be able to inspect, challenge, and seek independent review of uses concerning them. The arrangement remains sociotechnical (Orlikowski, 1992), while digital components' recombinability makes repurposing an ongoing governance concern (Yoo et al., 2010). Safeguards constrain but cannot eliminate power asymmetries.

The architecture supports, but does not produce, three forms of learning. *Single-loop learning* investigates signals and adjusts documentation, training, resources, or practice within existing *needs categories* and legal standards, then monitors needs- and rights-relevant evidence. *Double-loop learning* uses repeated signals or contestation to reconsider the categories, routines, assumptions, indicators, comparison rules, or safeguards shaping inquiry and action (Argyris & Schön, 1978). *Triple-loop learning* is used narrowly to ask who may authorize and redesign learning and accountability arrangements, and whose needs, rights, and knowledge define legitimate success. Given the term's diverse conceptualizations and limited empirical grounding, fuller theory and evaluation of all three forms remain future work (Tosey et al., 2012). *Informating* creates institutional visibility (Zuboff, 1988); *accountable digitality* additionally connects visibility to answerability, contestability, safeguards, authorized forums, response, and learning.

## *Evaluative Boundaries*

Legitimacy is not a system output. This paper distinguishes the normative question of whether authority accords with rights and justifiable standards from sociological legitimacy, understood as perceptions that institutional action is proper or appropriate (Suchman, 1995). Representation, perceived fairness and honesty, freedom from bias, correctability, and decision quality can shape evaluations of legal procedures (Tyler, 1988). *Needs-aware records* and protected inquiry may support relevant conditions—reason-giving, contestability, consistency review, and visible safeguards—but neither establish normative legitimacy nor guarantee perceived legitimacy, trust, or acceptance. Adverse effects such as surveillance, metric pressure, and decontextualized comparison also require empirical evaluation.

These mechanisms may enable accountability and learning but neither guarantee better outcomes nor exhaust possible realizations of the meta-requirements. Their effects depend on record quality, meaningful participation, lawful and proportionate access, privacy, professional practice, institutional capacity, and indicators remaining inquiry prompts rather than performance targets. They provide evaluable claims and boundary conditions without implying that selecting a candidate *satisfier* demonstrates *need satisfaction* or that an aggregate association proves bias or causation. Any consequential implementation would require context-specific evaluation and adaptation.

## Discussion and Implications for IS Research and Practice

### *Implications for IS Theory and Design*

Accountable digitality extends IS accountability research in three ways. First, it shifts the object of analysis from the *accountability of digital systems* to the *accountability effects of digital transformation* on institutions, consistent with socio-technical IS theory (Orlikowski, 1992; Yoo et al., 2010). Second, it identifies *needs-awareness* as a mechanism linking rights, data, and institutional learning: rights alone can be too general for audit, while data alone can be normatively thin. Third, it treats *auditability* as a design quality co-created through records, forms, analytics, AI assistance, professional routines, and oversight forums. The immediate illustrative case is child-welfare judgment, but the broader IS problem is how to make consequential human institutions accountable through digital transformation without fully automating their judgment. Across public administration, healthcare, education, and social services, confidential decisions are dispersed yet require system-level accountability (Rai, 2017a, 2017b). Accountable digitality therefore concerns whether digital infrastructures make institutional reasoning and disparities visible enough to explain and learn from them, not merely whether technology is adopted (Vial, 2019). This reorients digital transformation research toward digitally mediated institutional self-knowledge: what a court, agency, or hospital can know about its decision patterns, explain to affected persons, and learn without exposing private lives (Vial, 2019).

### *Why the Superlative "Best" Changes the Digital Transformation Obligation*

The child-welfare case has *distinctive normative force* because the governing language is the child's *"best* interests." This does not imply perfect outcomes or a duty to adopt every available technology. It makes feasible, rights-respecting procedural improvements relevant to institutional justification because *best*-interest decisions must identify relevant interests, explain their weighting, and show how the principle was applied (United Nations Committee on the Rights of the Child, 2013). The argument developed here is therefore an evaluative implication, not a free-standing legal conclusion: an institution should be able to explain why it did or did not examine a feasible procedure that could materially improve its capacity to identify, reason about, and protect children's interests while respecting other rights.

The procedural dimension is decisive. The Committee on the Rights of the Child describes the best-interest principle not only as a substantive right and interpretive principle but also as a rule of procedure (United Nations Committee on the Rights of the Child, 2013). Procedure is therefore part of the child's best interests. If a decision process systematically fails to see relevant *needs*, cannot learn from aggregate disparities, or cannot explain why comparable children receive radically different developmental contexts, the problem is not merely administrative. It concerns the procedure through which *best* interests become legally real.

*This point gives accountable digitality its strongest legal and ethical justification*. If a privacy-preserving and independence-preserving information system can improve *best*-interest assessment by revealing missing *needs*, detecting unexplained disparities, supporting child-rights assessment, and enabling institutional learning, the question is not simply whether digital transformation would be useful. The question is whether an institution can justify not evaluating such procedures. Children's rights can thereby transform accountable digitality from managerial modernization into a reasoned institutional obligation to evaluate a potential procedural improvement, not a predetermined duty to digitize or adopt.

Any resulting case for adoption is conditional and proportional. It is not an obligation to deploy a particular dashboard, AI model, or metric. Institutions should evaluate digital audit infrastructures when there is reason to believe they can improve *best*-interest procedures; implement them only under safeguards that protect children, families, confidentiality, participation, and judicial independence; and continuously review whether the system itself creates harms. The *superlative best* interests of the child strengthens reasons for improving procedure, while child rights constrain the means of improvement.

The argument is therefore neither technological determinism nor anti-digital caution. A court procedure that produces isolated reasons but cannot detect systemic disparities may be legally familiar, but familiarity is not a justification if better rights-respecting procedures are feasible. Conversely, a digital audit system that increases surveillance, exposes children, pressures judges, or substitutes metrics for judgment would fail the same best-interest test. Where accountable digitality improves the institution's capacity to protect children's *needs* and does so with fewer rights risks than the status quo, the institution has strong reasons to move toward it.

### *Implications for Courts and Child-Rights Governance*

For courts, the minimum stage should be evaluated through a single-site, non-decisional pilot rather than jurisdiction-wide deployment. A lawfully governed sample of closed or simulated cases would be manually represented as needs-aware reasoning records and examined only in protected aggregate form by an authorized multidisciplinary forum. Neither records nor aggregate outputs would alter individual judgments or rank judges, and AI would initially be excluded. Evaluation would examine feasibility, burden, missingness, reason quality, privacy, contestability, professional and affected-person acceptance, sociocultural adequacy, and whether each safe and significant parent receives equal procedural regard under the same needs- and evidence-based questions, subject to child-specific safety and other best-interest reasons, without presuming equal time. Document coding, interviews, privacy and accuracy tests, and process tracing would assess potential benefit, ritualization, gaming, and unintended effects (Orlikowski, 1992; Power, 1997). Optional AI would be considered only after a useful manual baseline. A later Austria-Sweden study could examine contextual transfer; it should not be treated as the initial pilot.

For child-rights governance, accountable digitality connects individual *best-interest assessment* with collective *child-rights assessment*. Aggregated data from individual decisions can inform whether the system as a whole respects children as rights-holders. Conversely, child-rights assessment can inform which needs and indicators should be documented in individual cases. This two-way connection is an important advantage of accountable digitality: it transforms confidential case-by-case decisions into protected institutional knowledge.

Engagement cannot be inferred from espoused goals of learning or betterment. Judges and other professionals should be able to see the system's lawful purpose, indicator definitions, uncertainty, comparison limits, access boundaries, and separation from individual performance ranking; children, parents, and representatives should be told what is recorded, who may see it, how it may be used, and how correction or contestation can be sought. Oversight bodies require method and data-lineage documentation sufficient for audit. Role-appropriate explanation, participation, contestation, non-punitive governance, and visible response to challenge may support warranted engagement, but trust and acceptance remain empirical outcomes and may decline if practice contradicts these commitments (Suchman, 1995; Tyler, 1988).

### *Implications Beyond Child Welfare*

Although the theory is developed through child-welfare judgment, that hard and paradigmatic running case is not the boundary of accountable digitality. In hiring, lawfully collected, purpose-limited monitoring data—kept separate from individual selection—can reveal persistent disparities across the applicant, shortlist, offer, and hire pipeline; such patterns should prompt authorized inquiry rather than constitute automatic proof of discrimination. Municipal *tree care* offers a more distant illustration focused on institutional practice and service output: sensor, work-order, and resource-allocation records can help distinguish exceptional weather from recurring watering gaps and examine whether scarce resources followed transparent urgency-based priorities. These are two illustrations among many, not coequal cases or evidence that the child-welfare architecture transfers unchanged. Each domain's purposes, needs or priorities, rights, lawful data uses, and safeguards must be derived anew.

The transferable design principle is needs-aware auditability. In each domain, the relevant needs, rights, risks, and safeguards will differ. The IS contribution is not a universal indicator set but a design grammar: identify the needs that justify the institution's authority; document how individual decisions link those needs to candidate satisfiers, evidence, reasons, uncertainty, and decisions; preserve privacy through layered visibility and data minimization; support aggregate audit; and ensure that detected patterns reach authorized human review, training, policy revision, or redress. This design grammar helps IS research connect *responsible digital transformation* to *institutional accountability* without presuming automated decision-making.

## Risks, Boundary Conditions, and Safeguards

A theory of *accountable digitality* must state its risks as clearly as its promise. The central risk is that an accountability infrastructure becomes a new source of harm: children become data objects, judges become metric targets, families experience intensified surveillance, and legal reasoning is narrowed to what the system can capture. These risks are not side issues. They are boundary conditions for the theory. *Accountable digitality* is normatively defensible only when digital transformation improves the institution's ability to protect needs and rights without undermining the confidentiality, dignity, participation, and independence that constrain and justify the procedure.

Table 1 assigns the principal architecture risks and safeguards to accountable actors and loci, but needs-aware audit systems can still fail. A contestable needs-and-candidate-satisfier vocabulary can become technocratic if disputed concepts are frozen; aggregate measures can become disciplinary tools; AI extraction can misread legal reasoning; and federated analytics retain privacy and governance risks. In bureaucratic settings, records and indicators can also produce unanticipated effects when formal compliance or a better score displaces the substantive purpose of protecting needs and rights (Merton, 1940; Power, 1997). Responses sometimes described as gaming are behaviorally ambiguous: boilerplate, strategic omission, or coding that obscures reasons is harmful manipulation, whereas adaptation that directs greater attention to *needs*, improves documentation, or increases challenge may be productive. Improvement cannot be inferred from better indicators in either case; qualitative inquiry and later needs- and rights-relevant evidence must be used, and categories and measures must remain contestable and periodically revisable (Espeland & Sauder, 2007; Selbst et al., 2019).

Two conditions bound responsible use of the architecture: institutional capacity and governance of public visibility. Without lawful authority, training, data stewardship, secure infrastructure, and a response procedure, audit may become symbolic or managerial. Public reporting may support answerability but also create reidentification, media pressure, and distorted comparisons; detailed or case-sensitive audit outputs should therefore be restricted to appropriate authorized actors and forums, while public reports should be highly aggregated and explicitly qualified (Nissenbaum, 2004; Power, 1997). Such visibility may support conditions relevant to procedural legitimacy, but it cannot guarantee legitimacy or trust.

## Conclusion

Digital accountability research shows that consequential digital systems should be documented, explainable, auditable, contestable, and governed (Ananny & Crawford, 2018; Kroll et al., 2017; Raji et al., 2020). This paper reverses the direction: digital transformation can make traditional institutions more accountable when designed as *accountable digitality*—a socio-technical arrangement that reveals institutional patterns while preserving the rights and normative commitments constraining and justifying institutional authority. The title names this reversal: *digital accountability* asks how digital systems are held to account; accountable digitality asks how becoming digital helps institutions account for themselves. Because consequential decisions remain human, confidential, and dispersed, neither refusing digital transformation nor delegating judgment to AI is sufficient. Rights-respecting information systems must make reasons and patterns knowable.

Child-welfare judgment is a hard and revealing case: best-interest reasoning and developmental knowledge must coexist with confidentiality and judicial independence. Needs-aware reasoning records, privacy-preserving aggregate audit, and institutional learning connect these elements. Generative or agentic AI may assist only as an accountable component that surfaces reasons and patterns rather than substitutes for judgment. The core claim is modest yet demanding: it promises neither automated objectivity nor a universal custody formula, but requires institutions affecting children to know and explain how judgments relate to children's needs across cases. Digital accountability helps IS research audit digital systems; accountable digitality can help society audit the institutions those systems increasingly mediate.

Future research should test the four contingent mechanisms with evidence that can support, refine, or challenge them. First, record coding, interviews, process tracing, and governed follow-up across professional and affected-person groups and linguistic and sociocultural settings can assess whether needs-aware reasoning records make needs, candidate satisfiers, evidence, alternatives, uncertainty, and reasons more contestable; ritualization, burden, silenced voice, cultural misrepresentation, or treating a selected satisfier as achieved need satisfaction would challenge the mechanism, while later evidence must separately assess need satisfaction. Second, retrospective data, simulations, privacy and accuracy audits, and contextual review can test whether protected omission or disparity signals survive case-mix, missingness, and coding-sensitivity checks; instability, group-patterned error, false inference, or unacceptable reidentification risk would challenge aggregation. Third, longitudinal observation, document analysis, and actor interviews can examine whether authorized forums produce contestation, correction, redress, or learning, and how mandate clarity, perceived control, professional norms, procedural fairness, resource demands, and role-appropriate explanation shape institutional and actor-level resistance, appropriation, acceptance, or disengagement; ritual review, punitive ranking, or ignored challenges would count against the mechanism (Kellogg et al., 2020; Orlikowski, 1992; Suchman, 1995; Tyler, 1988). Fourth, benchmark and human-factors tests and shadow-mode trials can assess whether templates or optional AI improve completeness or audit selection without substituting for judgment; automation bias, socioculturally or linguistically patterned error, surveillance, deskilling, or untraceable influence would challenge governed mediation. Fuller proposition theory can specify causal forms, moderators, and cross-domain boundaries after these mechanisms are evaluated.